\documentclass[twocolumn,showpacs,prb,floatfix]{revtex4}
\usepackage{epsfig}
\usepackage{tabularx}

\newcommand{\be}{\begin{equation}}
\newcommand{\ee}{\end{equation}}
\newcommand{\beqn}{\begin{eqnarray}}
\newcommand{\eeqn}{\end{eqnarray}}

\newcommand{\chisg}{\chi_{_{\mathrm{SG}}}}

\begin{document}

\title{Numerical studies of the two- and three-dimensional gauge 
glass\\at low temperature}
\author{Helmut G.~Katzgraber}
\affiliation{Department of Physics, University of
California, Davis, California 95616}

\date{\today}

\begin{abstract}
We report results from Monte Carlo simulations of the two- and
three-dimensional gauge glass at low temperature using 
parallel tempering Monte Carlo. In two dimensions, we find
strong evidence for a zero-temperature transition. By means of
finite-size scaling, we determine the stiffness exponent $\theta
= -0.39 \pm 0.03$. In three dimensions, where a
finite-temperature transition is well established, we find
$\theta = 0.27 \pm 0.01$, compatible with recent results from
domain-wall renormalization group studies.
\end{abstract}

\maketitle

\section{Introduction}
\label{intro}

The gauge glass is a model often used to describe
the vortex glass transition in high-temperature superconductors.
Still, there are areas which need to be understood. In two
dimensions, there is an ongoing controversy as to whether a
spin-glass transition occurs at finite temperature or not. In
three dimensions, a finite-temperature transition is well 
established\cite{olson:00}, although there is no consensus on the value 
of the stiffness exponent.

Recently, claims of evidence for a
finite-temperature transition in two dimensions with $T_c =
0.22$ by Kim\cite{kim:00} using resistively shunted
junction (RSJ) dynamics, and by Choi and
Park\cite{choi:99} who study the scaling of the spin-glass
susceptibility via Monte Carlo simulations, have been made.
In contrast, Granato\cite{granato:98} and Hyman {\em
et al.}\cite{hyman:95}, who also use RSJ dynamics find
evidence of a zero-$T$ transition. In addition, Monte Carlo
simulations by Fisher {\em et al.}\cite{fisher:91} and Reger and
Young\cite{reger:93} show evidence of a $T = 0$
transition, although the simulations were not performed at
low enough temperatures.

Here we perform Monte Carlo simulations of the two- and
three-dimensional gauge glass, using parallel
tempering\cite{hukushima:96,marinari:98b}, to go to
significantly lower temperatures than was possible in earlier
work. In particular, we are now able to cover the temperature
range in two dimensions where the claimed spin-glass
transition\cite{choi:99,kim:00} takes place. We find strong
evidence that $T_c=0$. 

In three dimensions, we study the gauge
glass at very low but finite temperatures to provide a good
estimate of the stiffness exponent $\theta$. Earlier estimates,
which were obtained from ground-state methods, are inconsistent.
One group\cite{reger:91,gingras:92,kosterlitz:97,maucourt:97}
finds values consistent with $\theta \approx 0$, whereas
another\cite{olson:00,akino:02,cieplak:92,moore:94,kosterlitz:98} finds
$\theta$ in the range $0.26$ -- $0.31$. We find that $\theta =
0.27 \pm 0.01$, which agrees with the results of
Refs.~\onlinecite{olson:00}, and 
\onlinecite{akino:02,cieplak:92,moore:94,kosterlitz:98}.

\section{Model, Observables, and Method}
\label{model}

The Hamiltonian of the gauge glass is given by
\begin{equation}
{\cal H} = -\sum_{\langle i, j\rangle} \cos(\phi_i - \phi_j - A_{ij}),
\label{hamiltonian}
\end{equation}
where the sum ranges over nearest neighbors on a hypercubic lattice
in $D$ dimensions of size $N = L^D$, and $\phi_i$ represent the
angles of the $XY$ spins. Periodic boundary conditions are
applied. The $A_{ij}$ are quenched random variables uniformly
distributed between $[0,2\pi]$\cite{katzgraber:01a}. Because
$A_{ij}$ represent the line integral of the vector potential
between sites $i$ and $j$, we have the constraint that $A_{ij} =
- A_{ji}$.

Traditionally, one uses the Binder ratio\cite{binder:81} to
estimate the critical temperature $T_c$. As for the gauge glass
the Binder ratio cannot exceed unity\cite{olson:00}, the
splaying of the curves is small, and $T_c$ is difficult to
establish. In order to avoid this problem, we use a method
introduced by Reger and Young\cite{reger:91} in which one
calculates the current $I$ defined as the derivative of the free
energy with respect to an infinitesimal twist to the
boundaries, i.e.,
\begin{equation}
I(L) = \frac{1}{L} \sum_i \sin(\phi_i - \phi_{i+\hat{x}} - 
A_{i\,i+\hat{x}}) \; .
\end{equation}
In this case, the twist is applied along the $\hat{x}$-direction.  
As $[\langle I(L)\rangle ]_{\rm av} = 0$, we actually calculate
the root-mean-square current $I_{\rm rms}$. Here 
$\langle \cdots \rangle$ represents a thermal average,
whereas $[\cdots]_{\rm av}$ represents a disorder average.
Note that the current scales\cite{katzgraber:02a} as
\begin{equation}
I_{\rm rms} = \tilde{I}(L^{1/\nu}(T - T_c)) \; ,
\label{scale_tgt0}
\end{equation}
for $T_c > 0$, whereas for $T_c = 0$,
\begin{equation}
I_{\rm rms} =L^{-1/\nu} \tilde{I}(L^{1/\nu}T) \; .
\label{scale_teq0}
\end{equation}
Equation (\ref{scale_teq0}) indicates that the $T=0$ stiffness
exponent $\theta$ is negative and equal to $-1/\nu$ (as
$I_{\rm rms} \sim L^\theta$).  Equation (\ref{scale_tgt0}) shows
that if $T_c >0$, the curves for $I_{\rm rms}$ for different
sizes {\em intersect}\/ at the critical point, whereas
Eq.~(\ref{scale_teq0}) shows that if $T_c = 0$ the data
{\em decrease}\/ with increasing size at $T=0$. For a
finite-temperature transition we expect $\theta > 0$, since then
the ordered state at $T=0$ is ``stiff'' on large scales, and so
will presumably resist small thermal fluctuations.  On the other
hand, for a zero-temperature transition, the system will then easily break up
under the influence of thermal fluctuations and therefore $\theta < 0$.

We also have calculated the spin-glass susceptibility, $\chisg$,
defined by
\begin{equation}
\chisg = N[\langle q^2\rangle]_{\rm av} \; ,
\label{chisg}
\end{equation}
where $q$ is the spin-glass order parameter:
\begin{equation}
q = \frac{1}{N} \sum_i^N \exp[i(\phi_i^\alpha - \phi_i^\beta)] \; .
\end{equation}
Here, $\alpha$, $\beta$ are two replicas of the system with the same
disorder. According to standard finite-size scaling the spin-glass
susceptibility,
defined in Eq.~(\ref{chisg}), behaves as
\begin{equation}
\chisg = L^{2 - \eta}\tilde{\chi}_{_{\mathrm SG}}(L^{1/\nu}(T - T_c))\; ,
\label{chi_scale}
\end{equation}
meaning that at criticality ($T = T_c$) it diverges with a power law,
i.e., $\chisg \sim L^{2 - \eta}$. Note that the power-law
prefactor in Eq.~(\ref{chi_scale}) with an unknown exponent
complicates the analysis of $\chisg$ compared with that for
$I_{\rm rms}$.

For the simulations, we use parallel tempering Monte 
Carlo\cite{hukushima:96,marinari:98b} as it allows us to study
systems at lower temperatures than with conventional methods.  
For details about the implementation, equilibration tests, and
detailed parameters of the simulations we refer the reader to
Ref.~\onlinecite{katzgraber:02a}. In two dimensions, the highest
temperature is $1.058$, whereas the lowest temperature is $0.13$
for $L = 4$, $6$, $8$, $12$, and $16$ (for $L = 24$ the lowest
temperature is $0.20$). The number of temperatures $N_T = 30$
is chosen to give satisfactory acceptance ratios for the Monte
Carlo moves between the temperatures ($N_T = 24$ for $L = 24$).  
In 3D the
lowest temperature studied is $0.05$ (note that $T_c
\approx 0.45$)\cite{olson:00} whereas the highest temperature
is $0.947$, and $N_T = 53$ for $L = 3$, $4$, $5$, $6$, and $8$.

\section{Results for $D = 2$}
\label{2dres}

\begin{figure}
\centerline{\epsfxsize=7.5cm \epsfbox{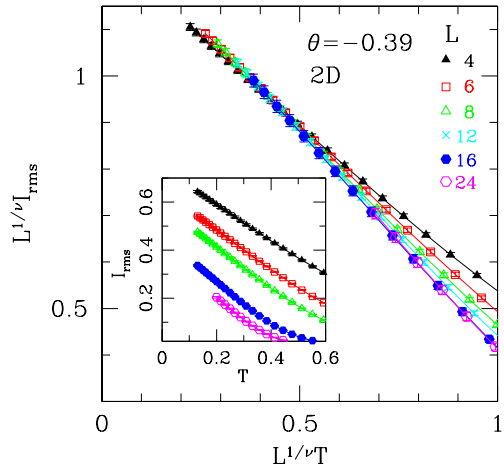}}
\vspace{-1.0cm}
\caption{
Scaling of the root-mean-square current $I_{\rm rms}$ in
two dimensions according to the form expected if $T_c=0$,
Eq.~(\ref{scale_teq0}). We see acceptable scaling of the data at
low temperatures. Deviations at higher $T$ are presumably due to
corrections to scaling. This plot is for $\theta \equiv -1/\nu =
-0.39$.
The inset shows $I_{\rm rms}$ as a function of $T$
for different system sizes. At all temperatures, the data decrease
with increasing $L$ indicating, from Eq.~(\ref{scale_tgt0}),
that if $T_c$ is finite it must be less than $0.13$.
}
\label{irms_scale}
\end{figure}

Figure \ref{irms_scale} shows a finite-size scaling plot of the
data
for $I_{\rm rms}$ in two dimensions
according to Eq.~(\ref{scale_teq0}). The
plot shows that the data collapse if $L^{1/\nu}T$ is small for
all values of $L$, and over the whole range of $L^{1/\nu}T$ for
the largest sizes.   
The inset shows results of the unscaled data.
At all temperatures, the data decrease with
increasing $L$ indicating, from Eq.~(\ref{scale_tgt0}), that
$T_c$ must be less than the range of temperatures studied.
The data in 
Fig.~\ref{irms_scale} are therefore consistent with a 
zero-temperature transition, but with significant corrections to
scaling at intermediate temperatures. We estimate the stiffness
exponent to be
$\theta = -0.39 \pm 0.03$.
The above error bar is estimated by varying $\theta$ slightly
until the data do not collapse well. This result is consistent
with recent work of Akino and Kosterlitz\cite{akino:02} who find
$\theta = -0.36 \pm 0.01$.
Choi and Park's\cite{choi:99} parameters
($T_c = 0.22$, $1/\nu = 0.88$) yield {\em very}\/ poor scaling,
especially near the proposed $T_c$. 
In fact, we are unable to get a reasonable fit to the data for 
$I_{\rm rms}$ according to Eq.~(\ref{scale_tgt0}) for 
{\em any}\/ finite $T_c$.

\begin{figure}[tb]
\centerline{\epsfxsize=7.5cm \epsfbox{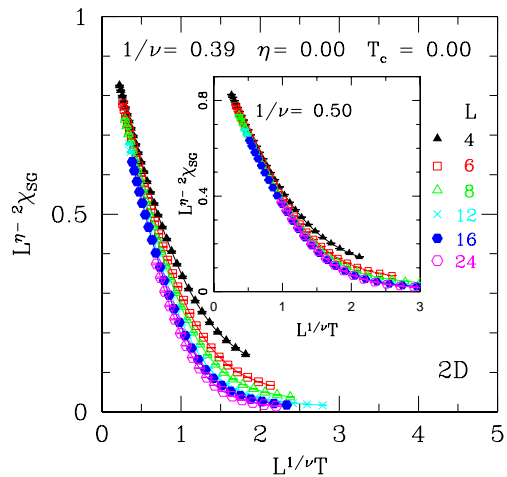}}
\vspace{-1.0cm}
\caption{
Scaling of $\chisg$ according to
Eq.~(\ref{chi_scale}) with $T_c = 0$. The data for large sizes
and low temperatures collapse with $\eta = 0$ and $1/\nu =
0.39$. The inset shows a scaling plot for the optimal value
$1/\nu = 0.50$ (and $\eta = 0$). For these values of exponents,
the collapse extends to a larger range of sizes.
}
\label{chisg_tc0}
\end{figure}

Figure \ref{chisg_tc0} shows a scaling plot of the 
data for $\chisg$ according to
Eq.~(\ref{chi_scale}) for $T_c = 0$ and $1/\nu = 0.39$, the
same parameters found in the scaling of $I_{\rm rms}$, together
with $\eta = 0.0 \pm 0.1$, which is expected at a zero-temperature
transition. The data at low $T$ and
for the largest sizes scale well, but the data away from this
range show deviations. Allowing $1/\nu$ to vary, we find $1/\nu =
0.50 \pm 0.03$.  The inset shows data for this optimal value,
where only the $L=4$ and $6$ data are not part of the scaling
function for all $L^{1/\nu} T$. The fact that the best values of
$1/\nu$ are not precisely the same when obtained from $\chisg$
and $I_{\rm rms}$ presumably indicates that scaling is only
valid for fairly low temperatures and large sizes, and that,
despite our working at quite low temperatures, we have only a
limited range of data which are fully in the scaling regime.

We also scale the data for $\chisg$ 
with Choi and Park's parameters\cite{choi:99}.
The data collapse is poor near $T_c$. The scaling
of $I_{\rm rms}$ is also much worse, indicating that $I_{\rm
rms}$ distinguishes between a finite $T_c$
and $T_c= 0$ much better than $\chisg$ because of its simpler finite-size
scaling form. An attempt to scale the data for $\chisg$ with
$T_c = 0.13$, the lowest temperature simulated, shows that the
best fit is obtained with $1/\nu = 0.68$ and $\eta = 0.19$.
While the data scale acceptably well, the data for $I_{rms}$
scale poorly.

According to Eq.~(\ref{chi_scale}) for
$T = T_c$, the data for $\chisg$
should lie on a straight line at $T_c$.
However, the data in the vicinity of $T = 0.22$, the transition
temperature claimed by Kim\cite{kim:00} and Choi and
Park,\cite{choi:99} show a strong downward curvature\cite{katzgraber:02a}
in our analysis,
indicating that this is actually {\em above}\/ $T_c$. Only
around the lowest temperature where we have data, $T = 0.13$, is
the curvature small, although it still greatly exceeds the error
bars. This indicates that $T_c < 0.13$, which is compatible with
our data for $I_{\rm rms}$.

\section{Results for $D = 3$}
\label{3dres}

Olson and Young\cite{olson:00} have investigated the critical
region of the three-dimensional gauge glass obtaining a lower
bound for the stiffness exponent of $\theta \ge 0.18$. Some
previous results\cite{reger:91,gingras:92,kosterlitz:97,maucourt:97} 
find $\theta$ in the range $0 \le \theta \le 0.077$ whereas
others\cite{akino:02,cieplak:92,moore:94,kosterlitz:98} find a
much larger value, $0.26 \le \theta \le 0.31$.

\begin{figure}[tb]
\centerline{\epsfxsize=7.5cm \epsfbox{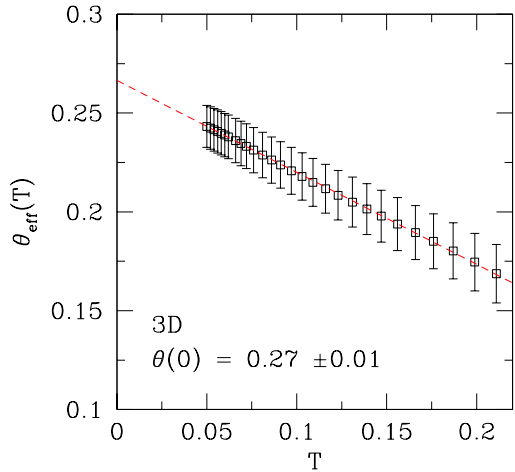}}
\vspace{-1.0cm}
\caption{
Effective stiffness exponent $\theta_{\rm eff}(T)$ as a function
of temperature in three dimensions for low temperatures. The
data extrapolate linearly to $T = 0$.
}
\label{theta}
\end{figure}

We can estimate $\theta$ from our data for $I_{\rm rms}$ since
$I_{\rm rms} \sim L^\theta$ when $L^{1/\nu}(T - T_c)$, the
argument of the scaling function in Eq.~(\ref{scale_tgt0}),
tends toward infinity. To obtain an estimate of $\theta$, we
perform a linear least-squares fit of $\ln (I_{\rm
rms})$ against $\ln (L)$ for each temperature in order to obtain an
{\em effective}\/ stiffness exponent $\theta_{\rm eff}(T)$ which
depends on the temperature. Figure~\ref{theta} shows that
$\theta_{\rm eff}(T)$ can be fitted well to a linear form at low
temperatures. Extrapolating to $T = 0$, we obtain
$\theta = 0.27 \pm 0.01$,
which is clearly positive and consistent with the results of
Refs.~\onlinecite{olson:00,akino:02,cieplak:92,moore:94,kosterlitz:98}.

\section{Conclusions}
\label{conclusions}

Our results from Monte Carlo simulations of the two-dimensional
gauge glass are consistent with a $T = 0$ transition
with a stiffness exponent $\theta = -0.39 \pm 0.03$. 
These results are incompatible with
the prediction made by Kim\cite{kim:00} and Choi and
Park\cite{choi:99} that $T_c = 0.22$.
In three dimensions we report the first reliable estimate of the
stiffness exponent from finite-temperature Monte Carlo
simulations. We find $\theta = 0.27 \pm 0.01$, which agrees with
the results of
Refs.~\onlinecite{olson:00,akino:02,cieplak:92,moore:94,kosterlitz:98}.

HGK acknowledges support from the National Science Foundation
under Grant No.~DMR 9985978 and would like to thank A.~P.~Young for
a fruitful collaboration.

\vspace*{-0.7cm}

\bibliography{refs}

\end{document}